\documentclass[fleqn,10pt]{wlscirep}

\usepackage{subfigure}
\begin{document}

\title{Basin entropy: a new tool to analyze uncertainty in dynamical systems}

\author[1,*]{Alvar~Daza}
\author[1]{Alexandre~Wagemakers}
\author[2]{Bertrand~Georgeot}
\author[3]{David~Gu\'ery-Odelin}
\author[1]{Miguel~A.F.~Sanju\'an}

\affil[1]{Nonlinear Dynamics, Chaos and Complex Systems Group, Departamento de  F\'isica, Universidad Rey Juan Carlos, M\'ostoles, Madrid, Tulip\'an s/n, 28933, Spain}

\affil[2]{Laboratoire de Physique Th\'eorique, IRSAMC, Universit\'e de Toulouse, CNRS, UPS, France}

\affil[3]{Laboratoire Collisions, Agr\'egats, R\'eactivit\'e, IRSAMC, Universit\'e de Toulouse, CNRS, UPS, France}

\affil[*]{alvar.daza@urjc.es}

\begin{abstract}

In nonlinear dynamics, basins of attraction link a given set of initial conditions to its corresponding final states. This notion appears in a broad range of applications where several outcomes are possible, which is a common situation in neuroscience, economy, astronomy, ecology and many other disciplines. Depending on the nature of the basins, prediction can be difficult even in systems that evolve under deterministic rules. 
From this respect, a proper classification of this unpredictability is clearly required. To address this issue, we introduce the basin entropy, a measure to quantify this uncertainty. Its application is illustrated with several paradigmatic examples that allow us to identify the ingredients that hinder the prediction of the final state. The basin entropy provides an efficient method to probe the behavior of a system when different parameters are varied.
Additionally, we  provide a sufficient condition for the existence of fractal basin boundaries: when the basin entropy of the boundaries is larger than $\log 2 $, the basin is fractal. 

\end{abstract}

\maketitle

\section{\label{sec:Introduction}Introduction}

Dynamical systems describe magnitudes evolving in time according to deterministic rules. These magnitudes evolve in time towards some asymptotic behavior depending on the initial conditions and on the specific choice of parameters. If a given dynamical system possesses only one attractor in a certain region of phase space, then for any initial condition its final destination is clearly determined. However, dynamical systems often present several attractors and, in these cases of multistability, elucidating which orbits tend to which attractor becomes a fundamental question. 

A basin of attraction \cite{nusse_basins_1996} is defined as the set of points that, taken as initial  conditions,  lead the system  to  a  specific  attractor. When there are two different attractors in a certain region of phase space,  two basins exist which are separated by a basin boundary. This basin boundary can be a smooth curve or can
be instead a fractal curve. The study of these basins can provide much information about the system since their topology is deeply related to the dynamical nature of the system. For example, systems with chaotic dynamics usually display basins of attraction with fractal structures \cite{aguirre_fractal_2009}.

The previous discussion applies typically to dissipative dynamical systems. However, for open Hamiltonian systems, where the concept of attractors or basins of attraction is meaningless, we can still define escape basins in an analogous way to the basins of attraction in a dissipative system. An escape basin, or exit basin, is the set of initial conditions that escapes through a certain exit. The H\'enon-Heiles Hamiltonian is a well-known model for an axisymmetrical galaxy and it has been used as a paradigm in Hamiltonian nonlinear dynamics.  It is a two-dimensional time-independent dynamical system, where orbits having an energy above the critical one can escape through one of the three different exits. It is widely known that when two or more escapes are possible in Hamiltonian systems, fractal boundaries typically appear \cite{aguirre_wada_2001}. 

\begin{figure*}
\begin{center}
\includegraphics[width=\textwidth]{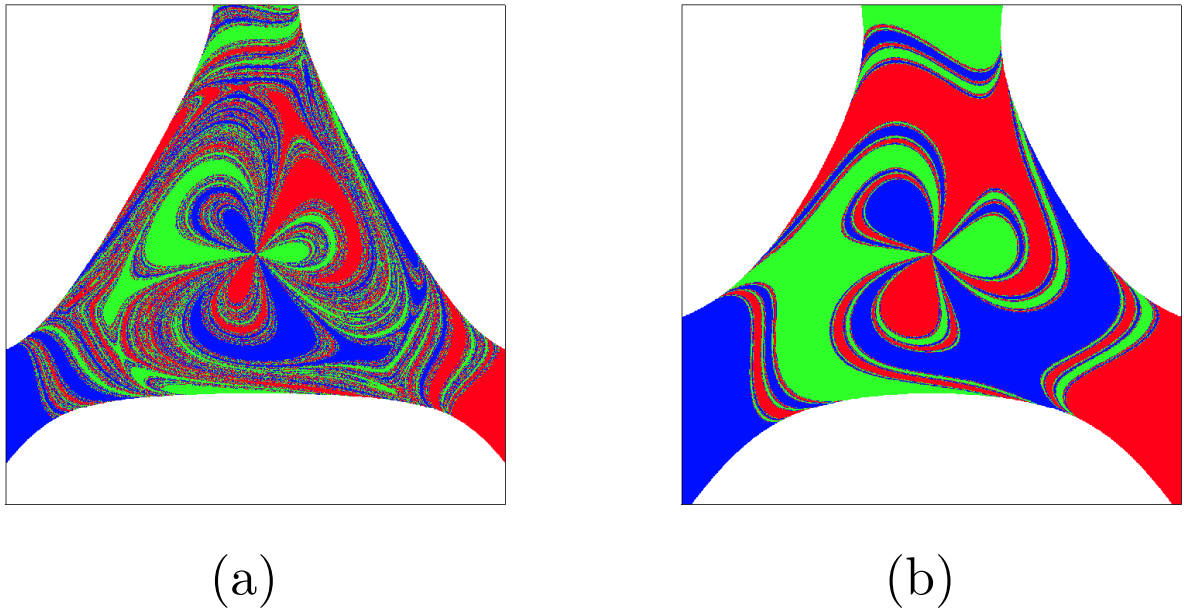}
\end{center}
\caption{\label{fig:AB} \textbf{Comparison between basins.} Escape basins for the H\'enon-Heiles system but different energies. They represent which exit will take each initial condition. It is clear that determining the final destination of the trajectories in the case (a) is harder than in the case (b). }
\end{figure*}

In order to give an intuitive picture of our problem we may look at Fig.~\ref{fig:AB}-(a) and Fig.~\ref{fig:AB}-(b). The figures show the escape basins of the H\'enon-Heiles Hamiltonian for two different values of the energy $E$ above the critical energy that separates bounded motions from unbounded motions. Most initial conditions leave the region through one of the three different exits to infinity for any $E$ above this critical energy. The colors represent points that taken as initial conditions leave the region through a specific exit. With this in mind, we may intuitively understand that it is harder to predict in advance which will be the final destination of an orbit in Fig.~\ref{fig:AB}-(a) than in Fig.~\ref{fig:AB}-(b).

The problem is that even though, we can have an intuitive notion that Fig.~\ref{fig:AB}-(a) is more uncertain than Fig.~\ref{fig:AB}-(b), there is no quantitative measure to affirm this. Moreover, this is not easy to assess when we compare two figures of basins corresponding to close values of the energy.

This is precisely the idea of uncertainty or unpredictability which we are considering here. This remark is important since we are aware that these terms are polysemic and consequently its use in the literature might be confusing.  In this paper we refer to unpredictability or uncertainty as the difficulty in the determination of the final state of a system, that is, to which attractor the initial conditions will tend to. Note that we speak about attractors for simplicity, though the discussion is identical for open Hamiltonian systems, where there are no attractors. This notion of unpredictability strongly differs from others used in nonlinear dynamics, like the Kolmogorov-Sinai entropy \cite{kolmogorov_new_1959,sinai_notion_1959}, the topological entropy \cite{adler_topological_1965}, or the expansion entropy \cite{hunt_defining_2015}, which refer to the difficulty of predicting the evolution of the trajectories. All these quantities are related to the topology of the trajectories, whereas our aim here is to develop an entropy depending on the topology of the basins.

The concept of basin of attraction is broadly used in all branches of science. The flow of water close to an obstacle can be described by means of basins of attraction, and their complicated structure explains the heterogeneity of phytoplankton and the information integration of the early macromolecules evolution \cite{karolyi_chaotic_2000}. Ideas traveling in a neuronal network can be expressed in terms of orbits moving among different basins of attraction \cite{rabinovich_transient_2008}. The decisions of agents subjected to changes in the market information exhibit complex dynamics, and this is reflected in their intricate basins of attraction \cite{brock_rational_1997}. The prediction of the evolution of interacting populations can be difficult when fractal boundaries separate the possible outcomes \cite{vandermeer_wada_2004}. These examples illustrate that we can gain much insight by measuring and understanding the uncertainty associated to the basins. 
 
Many authors describe fractal basin boundaries and when discussing its associated unpredictability, some vague affirmations are found due to a lack of an appropriated measure. In particular, this has been the case with the Wada basins which have received much attention in the past few years because they are said to be even more unpredictable than fractal basins \cite{kennedy_basins_1991,nusse_characterizing_2003,aguirre_unpredictable_2002,aguirre_wada_2001,vandermeer_wada_2004}. This affirmation appears repeatedly in the literature, and though it can be intuitively accepted, there is actually no quantitative basis for that.

Our paper constitutes an attempt to give a quantitative answer to the question of the uncertainty of the basins and this is precisely the problem that we discuss here. We propose a natural way to characterize the uncertainty of the basins by defining a quantitative measure that we call \textit{basin entropy}. The main idea is to build a grid in a given region of phase space, so that through this discretization a partition of the phase space is obtained where each element can be considered as a random variable with the attractors as possible outcomes. Applying the Gibbs entropy definition to that set results in a quantitative measure of the unpredictability associated to the basins. The discretization that we are considering arises naturally both in experiments and in numerical simulations. First, the experimental determination of initial conditions in phase space is physically impossible due to the intrinsic errors of the measurements. In the case of numerical experiments, the limitations of the computing resources constrain the resolution of the phase space under analysis. This unavoidable scaling error can induce wrong predictions even in deterministic models. Then, a natural question arises: how does the uncertainty in the initial condition affect the final state prediction?

\begin{figure*}
\begin{center}
\includegraphics[width=\textwidth]{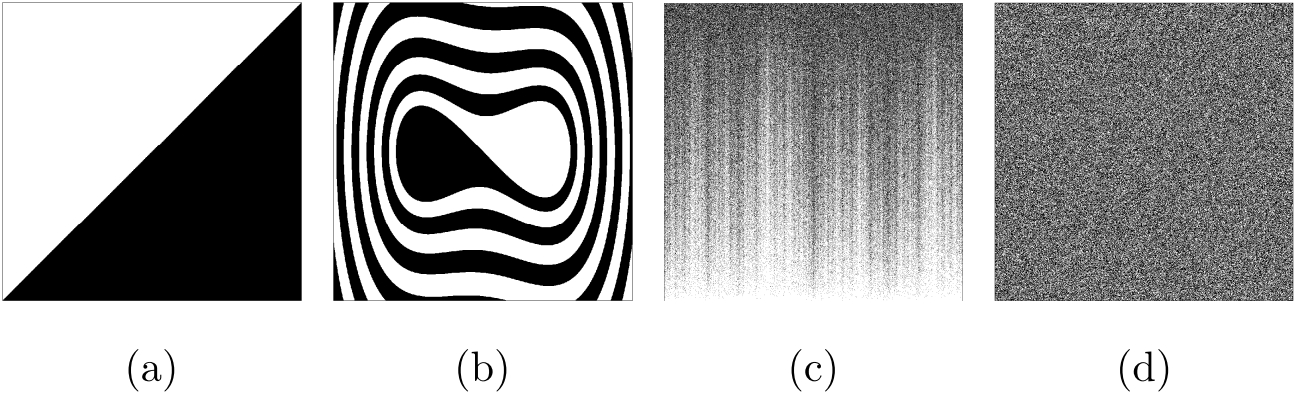}
\end{center}
\caption{\label{fig:method_comparison} {\bf Comparison of the different techniques.} The figure shows different basins obtained from well-known dynamical systems with two attractors. In panels (a) and (b), the uncertainty exponent is $\alpha=1$ since both boundaries are smooth, while in (c) and (d) $\alpha=0$ since both of them are riddled basins. The basin stability is equal to $1/2$ for the four basins. However, the basin entropy is able to distinguish the four cases and provides a method to measure quantitatively the unpredictability in increasing order from (a) to (d).}
\end{figure*}

A first approach to study the final state uncertainty has been investigated by Grebogi et al. \cite{grebogi_final_1983}. Given two attractors, they studied how the predictability of the system depends on the topology of their basins of attraction. They found a quantity $\alpha$ called uncertainty exponent, which is the dimension of the phase space $D$ minus the capacity dimension $d$ of the boundary that separates both basins
\begin{equation}
\alpha=D-d.
\label{eq:uncertainty_exponent}
\end{equation}
The uncertainty exponent takes the value $\alpha=1$ for basins with smooth boundaries, and $\alpha<1$ for basins with fractal boundaries. The closer $\alpha$ gets to zero the more difficult it becomes to predict the system. If smooth and fractal basins are mixed, the uncertainty exponent can still be calculated for each boundary, although the procedure is cumbersome \cite{grebogi_multi-dimensioned_1986}. As we will discuss later on, while the concept of uncertainty exponent is truly useful its application has several limitations.

Another approach to measure the unpredictability consists of evaluating the volume of each basin of attraction in a certain region of phase space. The ratio of the volume occupied by a single basin to the total volume defines the basin stability \cite{menck_how_2013}. It aims at classifying the different basins according to their relative sizes: larger basins are considered more stable. This notion has proved to be useful for the study of the stability of large networks of coupled oscillators, nevertheless it does not take into account how the basins are mixed. For different sets of parameters, a basin with two attractors can show smooth or fractal boundaries while the volume of each basin remains constant. The basin stability would be the same in both cases but obviously fractal boundaries have a more complex structure. A clear example is shown in Fig. \ref{fig:method_comparison}, where all the basins have the same basin stability. The uncertainty exponent also fails to capture the uncertainty associated to these basins. However, the basin entropy clearly distinguishes the four of them. In the following we provide the mathematical and computational foundation of the basin entropy and a method for its computation.

\section{\label{sec:Methods}Concept and definition of basin entropy}

Suppose we have a dynamical system with $N_A$ attractors for a choice of parameters in a certain region $\Omega$ of the phase space. We discretize $\Omega$ via a finite number of boxes covering it. Here we study two-dimensional phase spaces, so that we cover $\Omega$ with a grid of boxes of linear size $\varepsilon$. Now we build an application $C:\Omega\longrightarrow \mathbb{N}$ that relates each initial condition to its attractor, so that we will refer to that application as the \textit{color}. Each box contains in principle infinitely many trajectories, each one leading to a color labeled from 1 to $N_A$. In practice we can use only a finite number of trajectories per box. Indeed, it would correspond to the number of times an experiment is repeated, or the number of trajectories computed in a numerical simulation. In this work, we use square boxes with twenty-five trajectories per box (if not otherwise stated) in our numerical simulations. We have seen that twenty-five trajectories per box allows fast computation and provides accurate values of the basin entropy in all the cases studied here (see Fig.~S1 in the Supplementary Information) .

Although $\varepsilon$ is our limiting resolution, the information provided by the trajectories inside a box can be used to make hypotheses on the uncertainty associated to the box. We consider the colors into the box distributed at random according to some proportions. We can associate a probability to each color $j$ inside a box $i$ as $p_{i,j}$ which will be evaluated by computing statistics over the trajectories inside the box.

Taking into account that the trajectories inside a box are independent in a statistical sense, the Gibbs entropy of every box $i$ is given by

\begin{equation}
S_i=\sum\limits_{j=1}^{m_i} p_{i,j}  \log \left( \dfrac{1}{p_{i,j}}\right),
\end{equation}
where $m_i \in [1, N_A]$ is the number of colors inside the box $i$, and the probability $p_{i,j}$ of each color $j$ is determined simply by the number of trajectories leading to that color divided by the total number of trajectories in the box.

We choose non-overlapping boxes covering $\Omega$, so that the entropy of the whole grid is computed by the addition of the entropy associated to each one of the $N$ boxes of the grid 

\begin{equation}
S=\sum\limits_{i=1}^{N} S_i=\sum\limits_{i=1}^{N} \sum\limits_{j=1}^{m_i} p_{i,j} \log \left( \dfrac{1}{p_{i,j}}\right). \label{eq:entropy_definition}
\end{equation}
We note here that the growth of the number of boxes $N$ with the reduction of $\varepsilon$ provokes a counterintuitive effect: as we reduce the scaling box size $\varepsilon$ the entropy $S$ grows. In order to avoid this effect, we consider the entropy $S$ relative to the total number of boxes $N$ and define the following variable

\begin{equation}
S_b=\dfrac{S}{N},
\label{eq:Sb_definition}
\end{equation}
which we call \textit{basin entropy}. An interpretation of this quantity is associated to the degree of uncertainty of the basin, ranging from 0 (a sole attractor) to $\log N_A$ (completely randomized basins with $N_A$ equiprobable attractors).  This latter upper value is in practice seldom realized even for extremely chaotic systems. The basin entropy in general decreases with the scaling box size $\varepsilon$, as explained hereafter.
 We now have a tool to quantitatively compare different basins of attraction. 
 

Despite the fact that the basin entropy depends on the scaling box size $\varepsilon$, given a fixed $\varepsilon$ the value of the basin entropy converges as the number of trajectories inside a box increases.

At this point, we can delve deeper into the consequences of this definition by considering a simple hypothesis, which is to assume that the colors inside a box are equiprobable, thus $p_{i,j}=1/m_i, \forall j$. If we add the entropy of all the trajectories in a box, then  we recover the Boltzmann expression for the entropy $S_i= \log (m_i)$, where $m_i$ are the different colors inside a box (the accessible microstates of the Boltzmann entropy). Then the equiprobable total entropy becomes $S=\sum\limits_{i=1}^{N} S_i=\sum\limits_{i=1}^{N} \log (m_i).$ Furthermore, if we have a grid on a given region of phase space, many boxes will have an equal number of colors. That is, many boxes will be in the interior or lie near the boundary between two or more basins. Then we can say that there are $N_k$ equal boxes (in the sense that they have the same number of colors), where $k \in [1,k_{max}]$ is the label for the different boundaries. Boxes lying outside the basin boundaries do not contribute to the entropy as they only have one color. In other words, what matters is what happens at the basin boundaries. Then, the basin entropy reads 
\begin{equation}
S_b=\sum\limits_{k=1}^{k_{max}} \dfrac{N_k}{N} \log (m_k).
\label{eq:substituting}
\end{equation}
By following the method of the box-counting dimension $D_k$ \cite{alligood_chaos:_1996}, by which we compute fractal dimensions of basin boundaries, the number of boxes that contains a boundary grows like
$N_k = n_k \varepsilon ^{-D_k}$ where $n_k$ is a positive constant. In the case of smooth boundaries, the equation $D_k=D-1$ holds, $D$ being the dimension of the phase space. For fractal boundaries $D_k$ can be larger, but obviously we always have $D_k \leq D$. On the other hand, the number of boxes in the whole region of phase space, grows as $N = {\tilde n} \varepsilon^{-D},$ where $\tilde n$ is a positive constant. Substituting these expressions for $N_k$ and $N$ in Eq.~\ref{eq:substituting}, and recalling that $\alpha_k=D-D_k$ is the uncertainty exponent \cite{grebogi_final_1983} for each boundary, we get

\begin{equation}
S_b=\sum\limits_{k=1}^{k_{max}}  \dfrac{n_k}{\tilde n} \varepsilon^{\alpha_k} \log (m_k). \label{eq:3terms}
\end{equation}
This last expression reveals important information. The basin entropy has three components: the term $n_k/\tilde n$ is a normalization constant that accounts for the boundary size which is independent of $\varepsilon$; the term of the uncertainty exponent $\alpha_k$, is related with the fractality of the boundaries and contains the variation of the basin entropy with the box size; finally there is a term that depends on the number of different colors $m_k$. All these terms depend on the dynamics of the system, while the scaling box size $\varepsilon$ depends only on the geometry of the grid.

Equation~\ref{eq:3terms} sheds light into some interesting questions. First, we can compare smooth boundaries ($\alpha_k=1$) and fractal boundaries ($\alpha_k<1$). For both of them, smooth and fractal basins, we get $S_b \rightarrow 0$ when $\varepsilon \rightarrow 0$, but it converges faster in the smooth case. That is, it is more difficult for the basin entropy to decrease its value in a system with fractal boundaries. Despite other important factors, fractal boundaries introduce a larger uncertainty than the smooth ones. Furthermore, if $\alpha_k=0$ then $S_b>0$ no matter the scaling box size (this might happen in riddled basins \cite{alexander_riddled_1992,ott_scaling_1993,lai_geometric_1995}).

These ideas can be successfully applied for Wada basins. Basins exhibiting the Wada property have only one boundary that separates all the basins \cite{kennedy_basins_1991,daza_testing_2015}. We can argue that increasing the number of colors in the boundary boxes increases the basin entropy and therefore its uncertainty. In particular, having all possible colors in every boundary box is a unique situation found only in Wada basins. Nevertheless, Eq.~\ref{eq:3terms} also reveals that some non-Wada basins can show larger basin entropy than others exhibiting the Wada property. This can be the case when a system has the Wada property but there is one basin which occupies most of the phase space. Other factors like the number of attractors and the boundary size also play a role in the uncertainty according to the basin entropy formulation. Therefore the Wada property increases the uncertainty under the basin entropy perspective, but each case must be carefully studied.

The basin entropy idea can also be used to develop new tools. In some cases, we may be interested only in the uncertainty of the boundaries. In particular, we often want to know if a boundary is fractal. For that purpose we can restrict the calculation of the basin entropy to the boxes falling in the boundaries, that is, we can compute the entropy only for those boxes $N_b$ which contain more than one color,
\begin{equation}
S_{bb}=\dfrac{S}{N_b},
\end{equation}
where $S$ is calculated in the same way described before (see Eq. \ref{eq:entropy_definition}). We refer to this number $S_{bb}$ as \textit{boundary basin entropy}, because it quantifies the uncertainty referring only to the boundaries. 

The nature of this quantity $S_{bb}$ is different from the basin entropy $S_b$ defined in Eq.~\ref{eq:Sb_definition}. The $S_b$ is sensitive to the size of the basins, so it can distinguish between different basins with smooth boundaries, whilst the $S_{bb}$ cannot. However, it is worthwhile to introduce this new concept since it provides a sufficient condition to assess easily that some boundaries are fractal. Here is the reasoning. Suppose that we have several basins in a 2D phase space separated by smooth boundaries. Then, every box in the boundary will have only two colors, except a few countable number of boxes that may contain three colors or more. If we take a sufficient number of boxes in the boundaries, the effect of those boxes containing more than two colors will be negligible for the computation of the basin entropy in the boundaries $S_{bb}$. Then, the maximum possible value of $S_{bb}$ that a smooth boundary can show is $\log 2$, which would imply a pathological case where every box in the boundary contains equal proportions of two basins $p_i=1/2,\forall i\in \mathbb{N} $. Therefore, considering a sufficient number of boxes in the boundaries, we can affirm that if the boundary basin entropy is larger than $\log 2$, then the boundary is fractal, which can be expressed as

\begin{equation}
\label{log2}
S_{bb}>\log 2\Rightarrow \alpha<1.
\end{equation}
This is a sufficient but not necessary condition: as we shall discuss in Section~\ref{sec:map}, there may be fractal boundaries with  $S_{bb}<\log 2 $. Nevertheless, this threshold can be very useful to assess quickly the fractality of some boundaries, avoiding to compute the boundaries for different scales (which is not always possible). In Section~\ref{sec:map}, we will show on an example that the criterion (\ref{log2}) enables reliably to find parameter regions exhibiting fractal boundaries. A detailed proof of the $\log 2$ criterion can be found in the Supplementary Information.

\section{\label{sec:Results}What does the basin entropy measure?}

Here we illustrate the main features of basin entropy with several examples of dynamical systems, showing how its dependence on the boundary size $n_k/\tilde n$, the uncertainty exponent $\alpha_k$ and the number of attractors $N_A$. 

The term  $n_k/\tilde n$ corresponds to an estimate of the size of the boundary, since it normalizes the number of boxes containing the boundaries divided by the total number of boxes covering $\Omega$:

\begin{equation}
\dfrac{N_k}{N}=  \dfrac{n_k}{\tilde n} \varepsilon^{\alpha_k}.
\end{equation}

To study the contribution of this term, we consider the damped Duffing oscillator given by
\begin{equation}
\ddot{x}+\delta\dot{x}-x+x^3=0. \label{eq:damped_Duffing}
\end{equation}

\begin{figure*}
\begin{center}
\includegraphics[width=\textwidth]{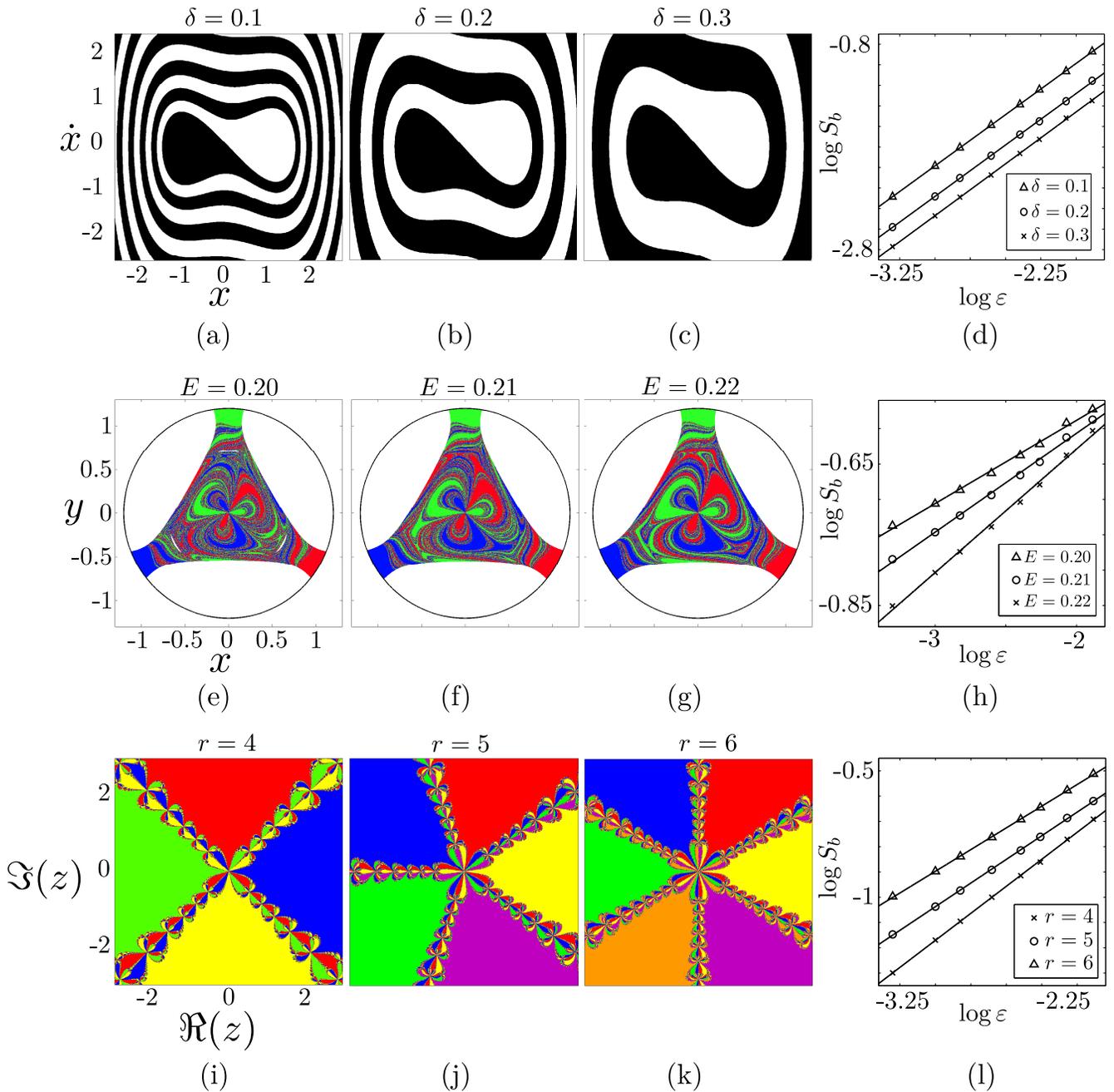}
\end{center}
\caption{\label{fig:panel} \textbf{Basin entropy ingredients.} (a)-(c) Basins of attraction of $\ddot{x}+\delta\dot{x}-x+x^3=0$. (d) Log-log plot of the basin entropy for different values of the scaling box size $\varepsilon$. The three basins have two attractors and $\alpha=1$ (same slope in the log-log plot), but the boundary size $n_k/\tilde n$ is different and the basin entropy reflects it for each case. (e)-(g) Escape basins of the H\'enon-Heiles Hamiltonian  $H=\frac{1}{2}(\dot{x}^2+\dot{y}^2)+\frac{1}{2}(x^2+y^2)+x^2y-\frac{1}{3}y^3$. (h) The log-log plot of the basin entropy shows different slopes for each case, since the uncertainty exponent $\alpha$ varies. (i)-(k)  The basins of attraction indicate the initial conditions that lead to the complex roots of unity using the Newton method described by $z_{n+1}=z_n-\dfrac{z^r-1}{r z^{r-1}}$. (l) The log-log plot shows that the basin entropy increases when the number of attractors increases, leading to larger values in the intercepts of the fits as predicted.}
\end{figure*}

This equation describes the motion of a unit mass particle in a double well potential with dissipation. This system presents two attractive fixed points in $(\pm1, 0)$ of the $(x, \dot{x})$ phase space. The higher the damping coefficient $\delta$ the faster the orbits tend to the fixed points and, as a consequence, the basin of attraction appears more deformed for smaller values of $\delta$ (Fig.~\ref{fig:panel}(a)-(c)). The damped Duffing oscillator is bistable, $N_A=2$, and has a smooth boundary with uncertainty exponent $\alpha=1$.

Observing the basins of attraction corresponding to the three different values of $\delta$, it is noticeable that the basin of Fig.~\ref{fig:panel}-(c) has a much simpler structure than the basin in Fig.~\ref{fig:panel}-(a). The outcome of an initial condition within an $\varepsilon$-box would be more difficult to predict in the second case. Nevertheless, both basins have the same uncertainty exponent $\alpha=1$ since in both cases the boundary is smooth. The differences in the values of the basin entropy originates from the differences in the region of discretized phase space occupied by the boundary, that is, the boundary size, which is reflected by the term $n/\tilde n$ (indices have been dropped since now there is only one boundary).

To highlight this effect, we have computed the basin entropy $S_b$ versus the scaling box size $\varepsilon$ for three different values of the damping coefficient $\delta$. The results are shown in the log-log plot of Fig.~\ref{fig:panel}-(d), where each fit corresponds to a different value of $\delta$. We must note that we have normalized the region of the phase space, so that the values of the scaling box size $\varepsilon$ in all the plots of the paper are the inverse of the number of pixels used as a grid. By taking logarithms on both sides of Eq.~\ref{eq:3terms}, we have

\begin{equation}
\log(S_b)= {\alpha} \log( \varepsilon) + \log \left(  \log (N_A)  \dfrac{n}{\tilde n} \right). \label{eq:fit}
\end{equation}

Since in this case, we have $\alpha=1$ and $N_A=2$ for all our simulations, it is clear that the variation of the basin entropy with $\delta$  is entirely due to the term $n/\tilde n$. Most importantly, we have obtained values of the slope $\alpha=1$ within the statistical error for all the fits. Therefore, although all these basins have the same uncertainty exponent, they have a different basin entropy for a given value of $\varepsilon$. The basin entropy is sensitive to their different structure and is able to quantify their associated unpredictability.

The fractal dimension of the boundaries also plays a crucial role in the formulation of the basin entropy. This is reflected in the uncertainty exponent $\alpha_k$ \cite{grebogi_final_1983} of Eq.~\ref{eq:3terms}. In order to highlight the effects of the variations in the uncertainty exponent, we have chosen a model that can display the Wada property \cite{aguirre_wada_2001}. This means that there is only one fractal boundary separating all the basins. The model is the H\'enon-Heiles Hamiltonian \cite{henon_applicability_1964},

\begin{equation}
H=\frac{1}{2}(\dot{x}^2+\dot{y}^2)+\frac{1}{2}(x^2+y^2)+x^2y-\frac{1}{3}y^3, \label{eq:HH}
\end{equation}
which describes the motion of a particle in an axisymmetrical potential well that for energy values above a critical one, the trajectories may escape from the bounded region inside the well and go on to infinity through three different exits.  If we vary the energy from $E=0.2$ to $E=0.22$, the fractal dimension of the boundaries is modified with $E$, though the Wada property is preserved \cite{blesa_escape_2012} (see Fig.~\ref{fig:panel}-(e)-(g)). The proportion of red, blue and green remains as a constant for these three basins, leading to constant values of the basin stability. However, the basin entropy accounts for their different structures.

As we compute the basin entropy for different scaling box sizes, we observe that the main effect of varying the parameter $E$ is a change of the slope in the log-log plot of Fig.~\ref{fig:panel}-(h). Equation~\ref{eq:fit} relates these changes in the slope to the uncertainty exponent $\alpha$ of the boundary. Smaller energies lead to smaller uncertainty exponents, since the boundaries have a more complex structure and consequently the slopes in the log-log plot decrease too. Obviously the offset also varies for the different values of the energy. This is related to changes in the boundary size $n/\tilde n$ which in this case cannot be completely separated from the changes in $\alpha$. This example shows that the scaling of the basin entropy with box size directly reflects the fractal dimension of the basin boundaries. For small box sizes this effect dominates and the largest fractal dimensions of the basins gives the largest basin entropies even though the offsets are different (see Fig.~\ref{fig:panel}-(h)).

The last factor that contributes to the basin entropy, according to Eq.~\ref{eq:3terms}, is the number of attractors $N_A$. In general, as the number of attractors increases, the uncertainty increases too, and so does the basin entropy.  Furthermore, it is impossible to isolate the effect of the number of attractors from the contribution of the boundary size, since they are not independent: if a new attractor emerges while tuning a parameter, a new boundary is also created. We illustrate these effects using a simple map where the number of attractors can be tuned. This map comes from the Newton method to find the complex roots of unity $z^r=1$ \cite{epureanu_fractal_1998}, and can be written as

\begin{equation}
z_{n+1}=z_n-\dfrac{z^r-1}{r z^{r-1}},\label{eq:newton}
\end{equation}
where $z\in \mathbb{C}$ and $r,n\in \mathbb{N}$. The attractors of this map are the solutions of $z^r=1$, so the parameter $r$ determines the number of attractors, $r=N_A$ (see Fig.~\ref{fig:panel}-(i)-(k) for $r=4,5,6$). The basins of attraction of this system have disconnected Wada boundaries, that is, all the basins share the same boundaries and are disconnected \cite{daza_testing_2015}.

From Eq.~\ref{eq:fit} we can predict that increasing the number of attractors increases the offset in the log-log plot of the basin entropy versus the box size. This can be observed in Fig.~\ref{fig:panel}-(l), where an increasing number of attractors leads to an increasing value of the basin entropy for all the $\varepsilon$ considered.

\section{Characterizing chaotic systems}

\subsection{\label{sec:map}Basin Entropy Parameter Set}

One of the most interesting applications of the basin entropy is to use it as a quantitative measure to compare different basins of attraction. We propose an analogy with the concept of \textit{chaotic parameter set} \cite{sanjuan_using_1998}, which is a plot that visually illustrates in a parameter plane when a dynamical system is chaotic or periodic by simply plotting the Lyapunov exponents for different pairs of parameters. Here, first we choose a given scaling box size $\varepsilon$, and then we evaluate the basin entropy associated to the corresponding basins of attraction for different parameter settings. We call the plot of the basin entropy in a two-dimensional parameter space \textit{basin entropy parameter set}. To illustrate the possibilities of this technique, we study the periodically driven Duffing oscillator $\ddot{x}+\delta\dot{x}-x+x^3=F \sin  \omega t$, whose dynamics can be very different depending on the parameters. We vary the forcing amplitude $F$ and the frequency $\omega$ of the driving, and for each basin we compute its corresponding basin entropy. We have used a resolution of $200$ $\times$ $200$ boxes ($\varepsilon=0.005$) with 25 trajectories per box (a million trajectories per basin) to compute the basins of attraction and the same region of the phase space $\Omega=[-2.5,2.5] \times [-2.5,2.5]$ for all the pairs $(F,\omega)$.

\begin{figure*}
\begin{center}
{\includegraphics[width=15cm]{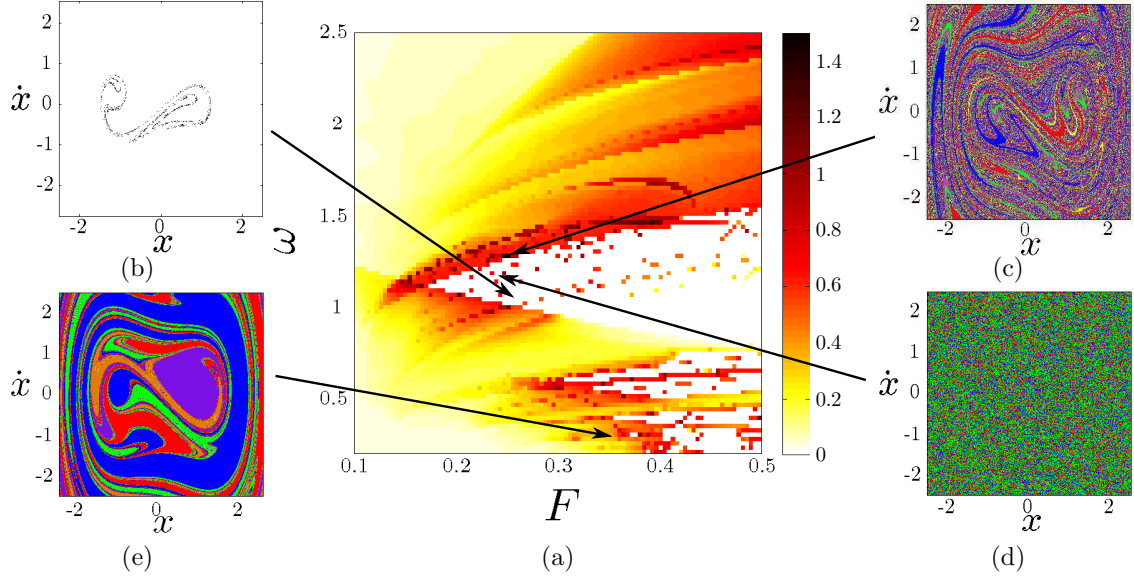}}
\end{center}
\caption{\label{fig:map} \textbf{Basin entropy parameter set.} (a) Basin entropy parameter set for the periodically driven Duffing oscillator given by $\ddot{x}+\delta\dot{x}-x+x^3=F \sin  \omega t $. It is a color-code map of the basin entropy for different values $(F,\omega)$ of the forcing amplitude and frequency, where we have fixed the scaling box size $\varepsilon=0.005$ and the damping coefficient $\delta=0.15$. We have used a color code where the \textit{hot} colors represent larger values of the basin entropy. (b) Example of a basin of attraction with zero basin entropy because there is only one attractor, actually a chaotic attractor (whose Poincar\'e section is plotted in black), for the parameters $F=0.2575$ and $\omega=1.075$. (c) Basins of attraction corresponding to the highest value of the basin entropy in this parameter plane, for $F=0.2495$ and $\omega=1.2687$. (d) Basins of attraction with three attractors and a very low uncertainty exponent happening for $F=0.2455$ and $\omega=1.1758$. (e) Basins of attraction with sixteen different attractors for the parameters $F=0.3384$ and $\omega=0.2929$.}
\end{figure*}

The result is presented in Fig.~\ref{fig:map}-(a), which is a color-code representation of the basin entropy in the  parameter plane $(F,\omega)$ for different values of the forcing amplitude and frequency. The \textit{hot} colors indicate higher values of the basin entropy, while the white pixels are for zero basin entropy. The set of parameters with zero basin entropy indicates that the basin of attraction has only one attractor. Although there is no uncertainty about the final attractor of any initial condition, trajectories may still be very complicated if the attractor is chaotic. This is actually the case for Fig.~\ref{fig:map}-(b), where there is only one chaotic attractor.

The \textit{hottest} point of the basin entropy parameter set corresponds to the basin of attraction shown in Fig.~\ref{fig:map}-(c) with eight different attractors whose basins are highly mixed. The reason for having this high value of the basin entropy lies at a combination of a high number of attractors and the uncertainty exponent associated to the boundaries that makes basins of attraction more unpredictable. In Fig.~\ref{fig:map}-(d), we can see a basin of attraction with extremely mixed basins, but it has only three attractors so its basin entropy is lower than for Fig.~\ref{fig:map}-(c). The converse situation arises in Fig.~\ref{fig:map}-(e), where there are sixteen different attractors but the boundaries are not very intricate.

Remarkably, it is also possible to explore the parameter space using only a few boxes instead of computing the high resolution basin for each parameter set. To infer a good approximation of the basin entropy, we applied a Monte Carlo sampling method. We have used 2000 boxes for each point in the parameter set, \textit{i.e.}, 50000 trajectories for each value of $(F,\omega)$,  instead of the million trajectories needed for the usual procedure (we mean by usual procedure computing the whole basin of attraction and then calculate the basin entropy). Thus, we speed up the computation by a factor 20. 

To evaluate the discrepancies between the usual procedure and the random sampling we have calculated the relative error $\varepsilon_{rel}=\frac{\vert S_b-S_b(RS)\vert}{\left\langle Sb \right\rangle } \times 100$. For the $94\%$ of the parameters evaluated the relative error in the basin entropy computation was less than $5\%$ (see Fig.~S2 in the Supplementary Information). If higher precision is desired one can always increase the number of boxes $N$, since the error decreases as $\frac{1}{\sqrt{N}}$ in the Monte Carlo method. Therefore, one can calculate the basin entropy using a small number of boxes and afterwards, one can compute with a finer grid the most interesting basins. The random sampling procedure is especially useful to compute the basin entropy for high dimensional systems or parameter sets.

\subsection{Log 2 criterion}

\begin{figure*}
\begin{center}
\includegraphics[width=\textwidth]{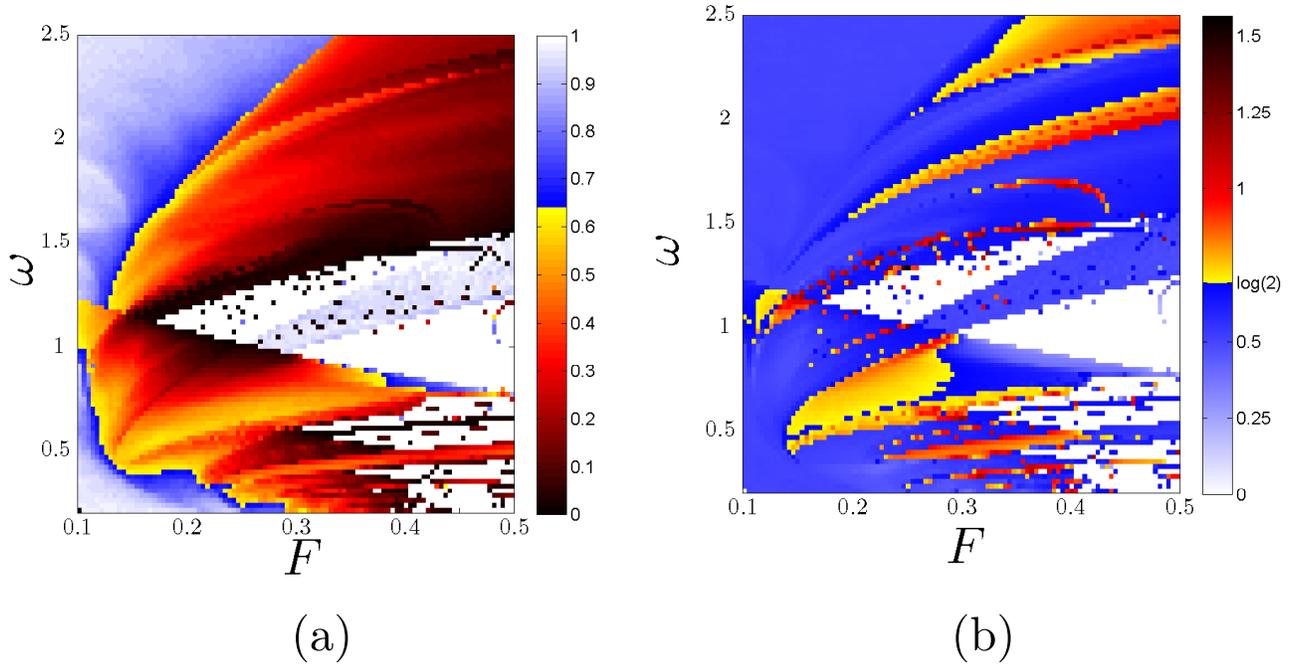}
\end{center}
\caption{\label{fig:Sbb} \textbf{Log 2 criterion comparison} (a) A color map of the boundary basin entropy $S_{bb}$ for different parameters $(F,\omega)$ for the periodically driven Duffing oscillator $\ddot{x}+\delta\dot{x}-x+x^3=F \sin \omega t$ for $\delta=0.15$ and $\varepsilon=0.005$. Hot colors are for basins with $S_{bb}>\log 2$ (b) Uncertainty exponent in the parameter plane. Hot colors indicate fractal boundaries. These figures confirm that the $\log 2$ criterion is a {\em sufficient but not necessary condition} for fractal boundaries.}
\end{figure*}

Using the same data, we can also study the boundary basin entropy $S_{bb}$ in the parameter plane. This quantity reflects the uncertainty associated to the boundaries, and we have seen in Section \ref{sec:Methods} that if $S_{bb}>\log 2$ then the boundary is fractal. For a given scaling box size $\varepsilon$, this process cannot distinguish a true fractal boundary from a smooth boundary which at this scale separates more than two basins inside one box. 
The results for the periodically driven Duffing oscillator are depicted in the colormap of Fig.~\ref{fig:Sbb}-(a), where white color is assigned to the pairs $(F,\omega)$ displaying only one attractor. By means of this plot we can detect parameter regimes where boundaries are fractal, depicted with hot colors. Figure~\ref{fig:Sbb}-(b) shows the values of the uncertainty exponent, using hot colors for the parameters with fractal boundaries. Comparing Fig.~\ref{fig:Sbb}-(a) and Fig.~\ref{fig:Sbb}-(b), we can confirm that the $\log 2$ criterion is a {\em sufficient but not necessary condition} for fractal boundaries. Indeed, the $\log 2$ criterion works only for cases with three or more basins (see Fig.~S3 in the Supplementary Information). Nevertheless the $\log 2$ criterion can be used to ascertain the fractality for basins of comparable size, and is much faster to compute than the direct determination of fractal dimension 
since it does not require the use of different scales. This makes it especially appealing for experimental settings where the resolution cannot be tuned at will.

\section{\label{sec:Discussion}Discussion}

The basin entropy quantifies the final state unpredictability of dynamical systems. It constitutes a new tool for the exploration of the uncertainty in nonlinear dynamics. This should become a very useful tool with a wide range of applications, as exemplified by the different systems that we have used to illustrate this concept. For instance, escape basins are widely used in astronomy, as shown in recent studies on the Pluto-Charon system \cite{zotos_orbit_2015}. In these investigations it is commonly argued that basins close to the escape energy present a \textit{higher degree of fractalization} \cite{zotos_escape_2015, ernst_fractal_2014}. Here we have shown an example of an open Hamiltonian system used in galactic dynamics, namely the H\'enon-Heiles potential, and we have been able to quantify its uncertainty for different values of the energy. 
 
Another kind of problems where basins of attraction are very common is in iterative algorithms. Such algorithms abound in all sort of research fields, where basins of attraction are used to visualize the sensitivity of different methods \cite{asenjo_visualizing_2013, van_turnhout_instabilities_2009}. In this work we have applied the basin entropy idea to a prototypical iterative algorithm: the Newton method to find complex roots. We have quantified the uncertainty associated to this algorithm for different numbers of roots. The basin entropy technique can be used to compare the performances of different algorithms or to see how modifications in some parameters like the damping may alter the uncertainty of the iterative processes. 

The concept of basin entropy also contributes to quantify the uncertainty of the Wada property, a recurring issue in the literature~\cite{kennedy_basins_1991,nusse_characterizing_2003,aguirre_unpredictable_2002,aguirre_wada_2001,vandermeer_wada_2004}. Moreover, using the idea of boundary basin entropy, we provide a sufficient condition to test the fractality of the boundaries. In contrast with other methods like the box-counting dimension that require computation at different resolutions, the $\log 2$ criterion can be used with a fixed resolution. We believe that this opens a new window for experimental demonstrations of fractal boundaries.

We have also proposed a new technique called basin entropy parameter set, that can flesh out the information given by bifurcation diagrams and chaotic parameter sets. Combined with Monte Carlo sampling, the basin entropy parameter set can also be used as a quick guide to find sets of parameters leading to simple or more complicated basins of attraction. 

We believe that the concept of basin entropy will become an important tool in complex systems studies with applications in multiple scientific fields especially those with multistability and other scientific areas as well.

\section*{Acknowledgments}

This work was supported by Spanish Ministry of Economy and Competitiveness under Project No. FIS2013-40653-P. Financial support from the Programme Investissements d'Avenir under the program ANR-11-IDEX-0002-02, reference ANR-10-LABX-0037-NEXT is also acknowledged.

\section*{Author contributions statement}

A.D., A.W., B.G., D.G.-O. and M.A.F.S. devised the research. A.D. performed the numerical simulations. A.D., A.W., B.G., D.G.-O. and M.A.F.S. analyzed the results and wrote the paper.

\section*{Additional information}

\textbf{Supplementary information} accompanies this paper.\\
\textbf{Competing financial interests}: The authors declare no competing financial interests.

\clearpage

\section*{Supplementary material}

\subsection*{Proof of the log 2 criterion}

The $\log 2$ criterion is a sufficient condition to prove the fractality of the basin boundaries. It is based on the concept of \textit{boundary basin entropy}, defined as
\begin{equation}
S_{bb}=\dfrac{S}{N_b},
\end{equation}
where $N_b$ is the number of boxes containing more than one color, that is, the number of boxes in the boundaries, and 
\begin{equation}
S=\sum\limits_{i=1}^{N} S_i=\sum\limits_{i=1}^{N} \sum\limits_{j=1}^{m_i} p_{i,j} \log \left( \dfrac{1}{p_{i,j}}\right). \label{eq:A3entropy_definition}
\end{equation}
Now we assume that the boundaries separating the basins are smooth. In this case, the number of boxes lying in the boundary separating two basins grows as
\begin{equation}
N_2=n_2\varepsilon^{-(D-1)},
\end{equation}
where $D$ is the dimension of the phase space. For $D=2$, the boundary would be a line, for $D=3$, it would be a surface and so forth. However, there might be some boxes $N_k$ lying in the boundaries of $k>2$ different basins. These boxes are in the intersection of at least two subspaces of dimension $D-1$, that is, they are in the intersection of two smooth boundaries. For instance, when $D=2$, it simply means that two or more smooth curves intersect in a point or collection of points, and when $D=3$, two or more smooth surfaces intersect forming smooth curves. Thus, the dimension of the subspace separating more than two basins must be $D-2$, and the boxes $N_k$ belonging to this subspace must grow as
\begin{equation}
N_k=n_k\varepsilon^{-(D-2)}.
\end{equation}
Taking into account that the total number of boxes grows as $N=\tilde n\varepsilon^{-D}$, we can express $N_2$ in terms of $N$ as
\begin{equation}
\label{eq:N2}
N_2=n_2 \left( \dfrac{N}{\tilde n}\right) ^{\frac{D-1}{D}},
\end{equation}
and for the boundary boxes separating more than two basins $N_k$, we have
\begin{equation}
\label{eq:N3}
N_k=n_k \left( \dfrac{N}{\tilde n}\right) ^{\frac{D-2}{D}}.
\end{equation}

At this point, we recall that the maximum possible value of $S$ in a box with $m$ different colors is $S=\log m$, which is the Boltzmann expression for the entropy of $m$ equiprobable microstates. Then, we can find that all the boxes in the boundary of two basins have $S\leqslant\log 2$, while for boxes in the boundary of $k$ basins, $k>2$, we have that $S\leqslant\log k$. Notice that the equality of the previous equations would be possible only in a pathological case where all the boxes in the boundaries have equal proportions of the different colors.

Then, the basin entropy $S_{bb}$ for this hypothetical system with smooth boundaries is
\begin{equation}
S_{bb}\leq\dfrac{N_2\log 2+N_k\log k}{N_2+N_k}.
\end{equation}
By substituting $N_2$ and $N_k$ by Eqs. \ref{eq:N2}-\ref{eq:N3}, we obtain the following expression
\begin{equation}
S_{bb}\leq\dfrac{n_2 \left( \dfrac{N}{\tilde n}\right)^{\frac{D-1}{D}} \log 2+n_k  \left( \dfrac{N}{\tilde n}\right)^{\frac{D-2}{D}} \log k }{n_2  \left( \dfrac{N}{\tilde n}\right)^{\frac{D-1}{D}}+n_k  \left( \dfrac{N}{\tilde n}\right)^{\frac{D-2}{D}}},
\end{equation}
which can be simplified as
\begin{equation}
S_{bb}\leq\dfrac{n_2 N\log 2+n_k \tilde n\log k}{n_2 N+n_k \tilde n},
\end{equation}
where $\tilde n,n_2,n_k$ are constants. Finally, we can take the limit of the previous inequality for a large number of boxes, that is when $N\rightarrow\infty$, leading to
\begin{equation}
\label{eq:ineq}
\lim_{N \to \infty}S_{bb}\leq\log 2.
\end{equation}
Therefore, we have proven that if the boundaries are smooth, then $S_{bb}\leq\log 2$, which is the same as to say that if $S_{bb}>\log 2$, then the boundaries are not smooth, i.e., they are fractal. This is what we call the $\log 2$ criterion.

This criterion is especially useful for experimental situations where the resolution cannot be arbitrarily chosen. In these cases we have a fixed value $\varepsilon>0$. 
Nevertheless, if we take a sufficient large number of boxes $N$, then the $\log 2$ criterion holds. Moreover, the equality of Eq.~\ref{eq:ineq} never takes place, so that there is some room for the possible deviations caused by the impossibility of making an infinite number of simulations or experiments.

\begin{figure*}
\begin{center}
{\includegraphics[width=8cm]{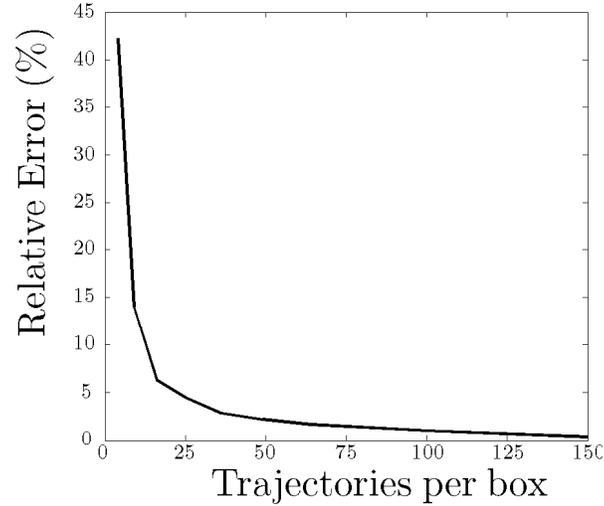}}
\end{center}
\caption{\label{fig:error_boxes} \textbf{Convergence of the basin entropy with the number of trajectories per box.} This figure represents the relative error in computing the basin entropy of Fig.~3-(d), taking as a reference the computation made with 2500 trajectories per box. In spite of the particularities of a given dynamical system (for instance the number of attractors is an important factor), we have seen that choosing 25 trajectories per box keeps the relative error below $5\%$ in most cases and allows a fast computation. Therefore, the number of trajectories per box is a parameter that can be tuned in order to get accurate results in a short time.}
\end{figure*}

\begin{figure*}
\begin{center}
\subfigure[~]{\includegraphics[width=8cm]{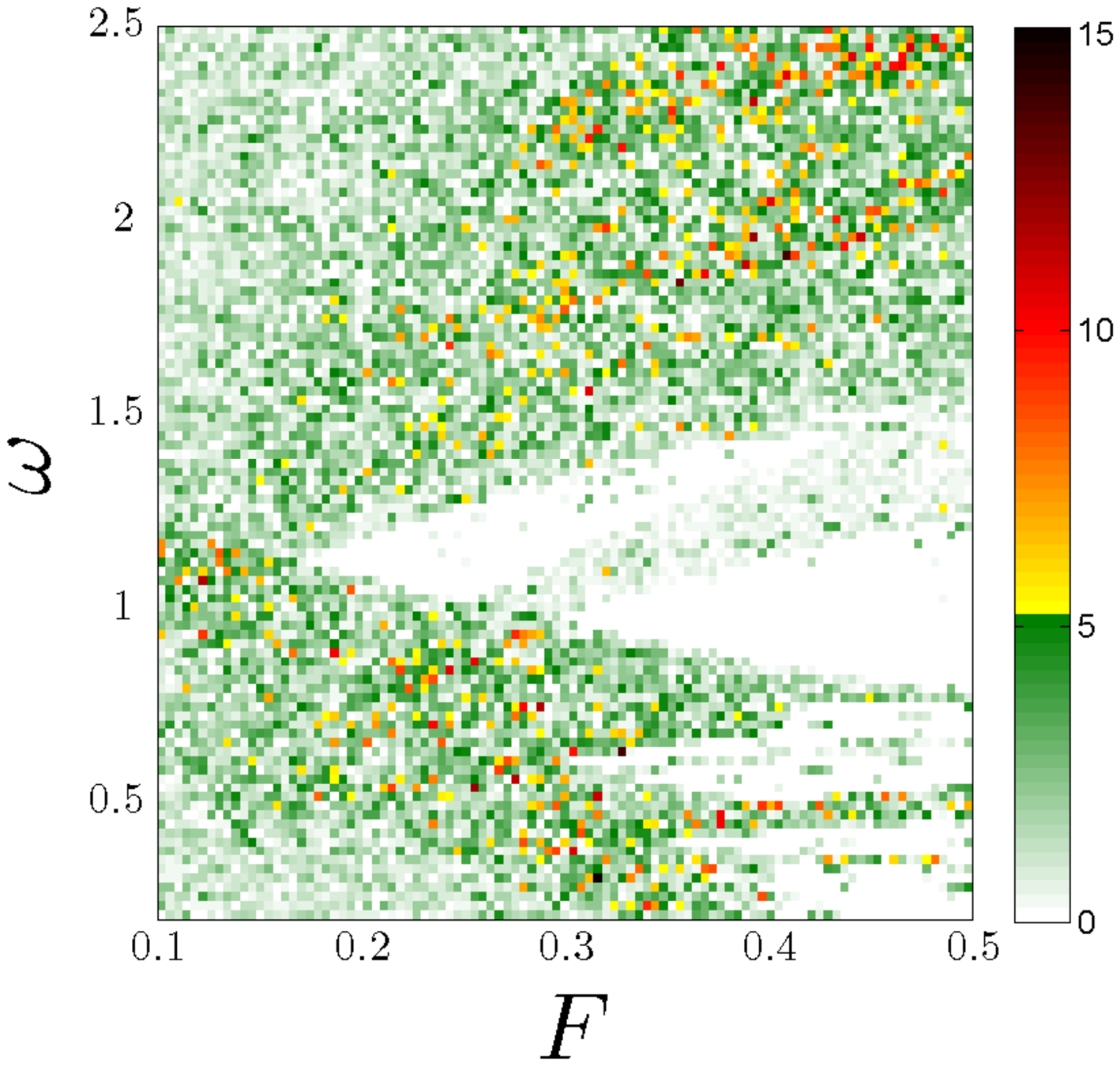}}
\subfigure[~]{\includegraphics[width=8cm]{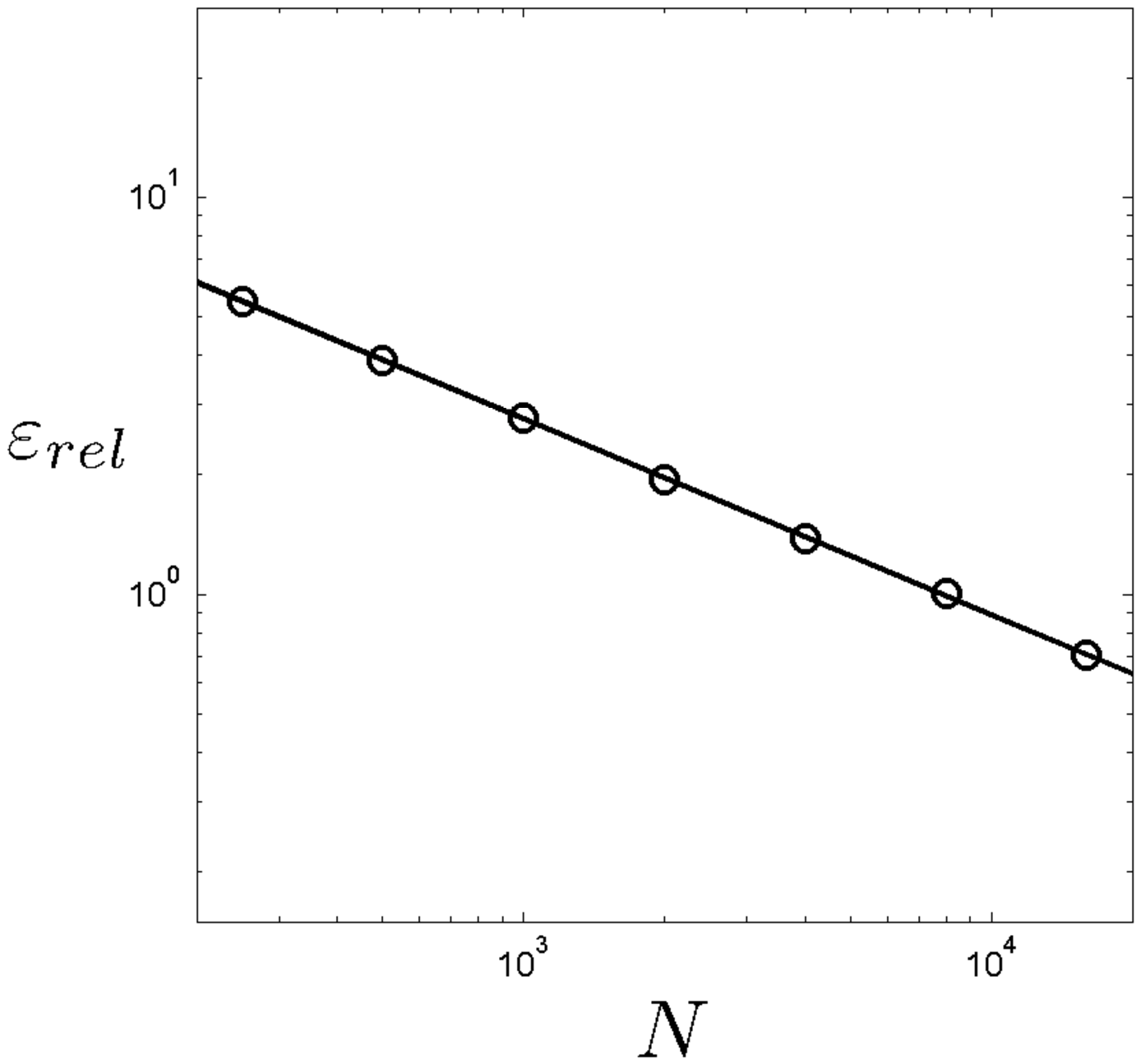}}
\end{center}
\caption{\label{fig:error} \textbf{Random sampling error.}  (a) This plot presents the relative error of the basin entropy estimation, that is $\frac{\vert S_b-S_b(RS)\vert}{\left\langle Sb \right\rangle } \times 100$, using 2000 boxes for the random sampling. The $94\%$ of the times, the relative error is below $5\%$. (b) If a more precise value is needed this error decreases as $\frac{1}{\sqrt{N}}$.}
\end{figure*}

\begin{figure*}
\begin{center}
\subfigure[~]{\includegraphics[width=8cm]{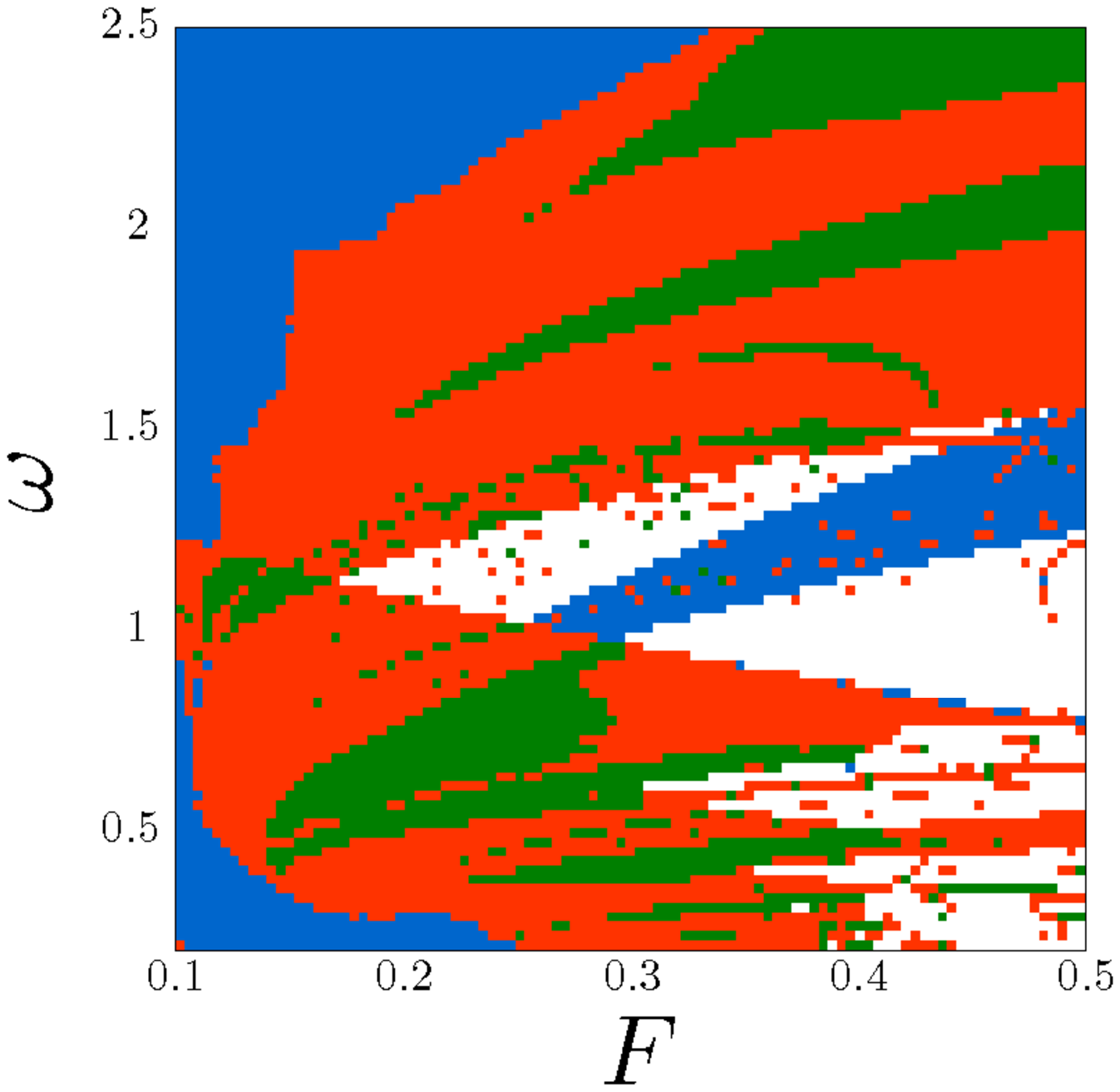}}
\subfigure[~]{\includegraphics[width=8cm]{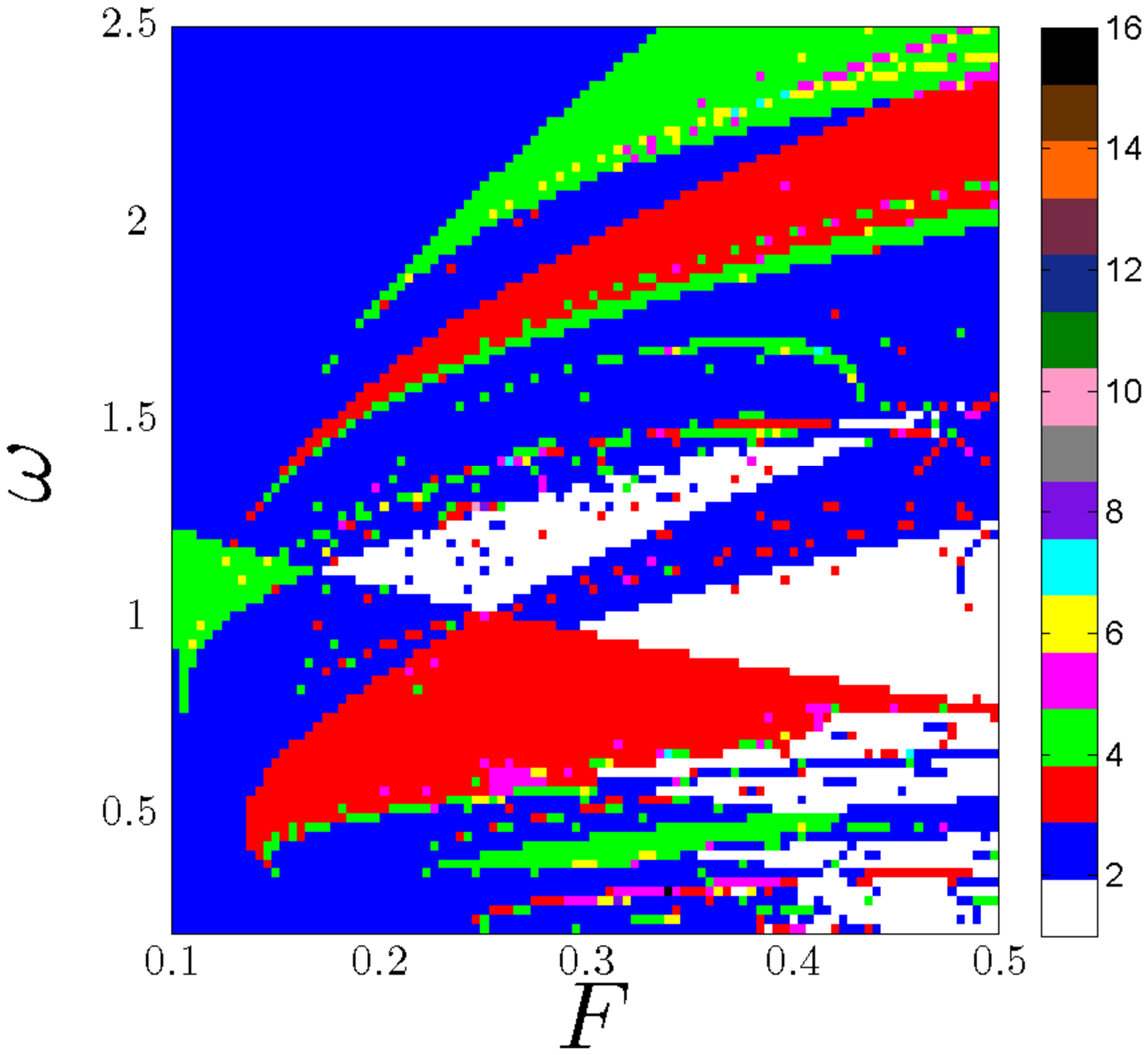}}
\end{center}
\caption{ \textbf{Some limitations to the log 2 criterion.} (a) White pixels indicate basins with one attractor, blue is for smooth boundaries, orange for fractal boundaries and green for fractal boundaries with $S_{bb}>\log 2$. All the basins with $S_{bb}>\log 2$ are fractal, but not all the fractal basins have  $S_{bb}>\log 2$. The $\log 2$ criterion is a \textit{sufficient but not necessary} condition for fractal boundaries. (b) Number of attractors in the parameter plane. The $\log 2$ criterion can only be fulfilled for basins with three or more attractors. }
\end{figure*}

\end{document}